\documentclass[aps,preprint,onecolumn ,amsmath,amssymb,showpacs,prb,unsortedaddress]{revtex4-1} 
\usepackage{epsf}      
\usepackage{graphicx}
\usepackage{gensymb}
\usepackage{sidecap}

\makeatletter

\@ifundefined{textcolor}{}
{
 \definecolor{BLACK}{gray}{0}
 \definecolor{WHITE}{gray}{1}
 \definecolor{RED}{rgb}{1,0,0}
 \definecolor{GREEN}{rgb}{0,1,0}
 \definecolor{BLUE}{rgb}{0,0,1}
 \definecolor{CYAN}{cmyk}{1,0,0,0}
 \definecolor{MAGENTA}{cmyk}{0,1,0,0}
 \definecolor{YELLOW}{cmyk}{0,0,1,0}
}

\usepackage{color}\usepackage{url}\usepackage{epsf}

\begin{document}

\title {Angle-Resolved Photoemission Spectroscopy Study of BaCo$_2$As$_2$
%Electronic structure of BaCo$_2$As$_2$: An angle-resolved photoemission study
%and full-potential linearized augmented-plane-wave calculations
}

\author{R. S. Dhaka}
\email{rsdhaka@ameslab.gov}
\author{Y. Lee}
\author{V. K. Anand}
\author{D. C. Johnston}
\author{B. N. Harmon}
\author{Adam Kaminski}
\email{kaminski@ameslab.gov}
\affiliation{Ames Laboratory, U. S. DOE and Department of Physics and Astronomy, Iowa State University, Ames, Iowa 50011, USA}

\date{\today}                                

\begin{abstract}
 We use angle-resolved photoemission spectroscopy and full-potential linearized augmented-plane-wave (FP-LAPW) calculations to study the electronic structure of BaCo$_2$As$_2$. The Fermi surface (FS) maps and the corresponding band dispersion data (at  90~K and 200~K) reveal a small electron pocket at the center and a large electron pocket at the corner of the Brillouin zone. Therefore the nesting between electron and hole FS pockets is absent in this compound, in contrast to the parent compounds of FeAs-based high-{\it T}$_{\rm c}$ superconductors.  The electronic structure at about 500~meV binding energy is very similar to features at the chemical potential in BaFe$_2$As$_2$. This indicates that complete substitution of Co for Fe causes a nearly rigid shift in the Fermi energy by adding two electrons per formula unit without significant changes in the band dispersions. The experimental FS topology as well as band dispersion data are in reasonable agreement with the FP-LAPW calculations. 
\end{abstract}

\maketitle

\section{Introduction}
Iron-arsenic-based high-temperature superconductors \cite{Kamihara08,Takahashi08,XHChen08,TYChen08,GFChen08,Rotter08,Yuan09,Cruz08,SefatPRL08} encompass a large number of materials with remarkably similar phase diagrams, despite significant differences in their chemical compositions. Both elemental substitutions and external parameters can be used to tune their physical properties.\cite{Takahashi08,XHChen08,TYChen08,GFChen08,Rotter08,Yuan09,Cruz08,SefatPRL08,Johnston10,CanfieldRev10} For example, in BaFe$_2$As$_2$ there is a high-temperature tetragonal to low-temperature orthorhombic structural phase transition at $\approx135$~K accompanied by an antiferromagnetic spin density wave (SDW) transition. The SDW ground state can be suppressed in a number of ways, which in many cases leads to an emergence of superconductivity.\cite{Takahashi08, XHChen08, TYChen08, GFChen08, Rotter08, Yuan09, Cruz08, SefatPRL08, Johnston10, CanfieldRev10} These Fe-based superconductors  crystallize in the ThCr$_2$Si$_2$-type structure that consists of a square lattice of iron ions sandwiched by pnictogen or chalcogen layers forming tetrahedral chemical bonds with Fe. Iron Fe$^{2+}$ ions  play an important role in the supercondcutivity, as the main contribution to the density of states near the Fermi energy is from the Fe $d$ orbitals.\cite{SinghPRL08} The Fermi surface (FS) of BaFe$_2$As$_2$ consists of hole and electron pockets at the center ($\Gamma$-point) and at the corner ($X$-point) of the Brillouin zone (BZ), respectively, of the body-centered-tetragonal direct lattice containing two formula unit (f.u.). There are hole and electron pockets that have very similar diameters, which leads to nesting that is most likely responsible for the SDW ordering\cite{Mazin10,MazinPRL08} and antiferromagnetism in the {\it A}Fe$_2$As$_2$ ({\it A} = Ca, Sr, Ba, Eu) parent compounds.\cite{Cruz08,Mazin10,Dhaka12} Previous studies demonstrated that an antiferromagnetic SDW phase is the ground state of these materials and leads to a substantial reconstruction of the Fermi surface at low temperature,\cite{LiuPRB09,YangPRL09,ZhangPRL09,Hsieh,KondoPRB10,LiuPRL09,YiPRB09} which disappears close to the carrier concentration at which the onset of the superconducting dome begins.\cite{LiuNP10,DhakaPRL11} The Co-substituted BaFe$_2$As$_2$ compounds are of particular interest since a rigid-band-like change of the chemical potential is reported with Co concentration.\cite{LiuNP10, LiuPRB11}. It has been shown, using angle-resolved photoemission spectroscopy (ARPES), that the Co substitutions induces a Lifshitz transition at the onset of the superconducting dome\cite{LiuNP10} and another change in the FS topology occurs with an electron pocket appearing around the $\Gamma$ point at higher ($x\ge0.24$) substitution levels,\cite{LiuPRB11} which is due to the change in the chemical potential that occurs by providing extra $d$ electrons to the system.

The large variation of properties of iron arsenic superconductors upon elemental substitutions calls for systematic investigations of the closely related materials, where all Fe atoms are substituted by other transition metals. Knowledge of their properties and electronic structure will not only improve understanding of the iron arsenic superconductors, but may also result in discoveries of novel materials with unusual and useful properties. 
%\textcolor{red}{
In particular, the success in synthesizing Co-based non-superconducting materials {\it A}Co$_2$As$_2$ ({\it A} = Ca, Sr, Ba, Eu) have attracted significant attention because of their many interesting physical properties.\cite{Ying12,Cheng12,Bishop10,Sefat09} BaCo$_2$As$_2$ exhibits paramagnetic behavior where the absence of magnetic ordering is considered to be caused by quantum fluctuations.\cite{Sefat09} Indeed, this compound is suggested to be a highly mass-renormalized metal near a ferromagnetic quantum critical point.\cite{Sefat09} It is believed that antiferromagnetic spin fluctuations associated with a SDW are the pairing glue for superconductivity in the iron arsenides. However, ferromagnetic spin fluctuations are strongly pair breaking for spin-singlet pairing. The calculated electronic density of states (DOS) of BaCo$_2$As$_2$ at  $E_{\rm F}$ is similar to those of the FeAs-based compounds; however, the extra $d$ electron in Co$^{2+}$ relative to Fe$^{2+}$ shifts $E_{\rm F}$ upwards in energy.\cite{Sefat09} This leads to the presence of a sharp peak in the DOS at  $E_{\rm F}$, which is derived from the Co $d$ orbitals.\cite{Sefat09} Therefore, the electronic structure of BaCo$_2$As$_2$ appears very different at  $E_{\rm F}$ than for the {\it A}Fe$_2$As$_2$ compounds. %A wide variety of physical phenomena of crystalline materials, such as, transport, optical and magnetic response, and phase transitions, rely on details of the topology of the FS. 
It is of great interest to clarify the band structure of BaCo$_2$As$_2$ near $E_{\rm F}$, which can provide important insights into the microscopic mechanism of the peculiar magnetic and transport properties as of the {\it A}Co$_2$As$_2$ compounds. So far there are very few studies of this very interesting material \cite{Ding}. 

In this paper, we present ARPES measurements on single crystals of BaCo$_2$As$_2$\cite{Pfisterer} that were carried out to clarify the above issues. We also present LDA band structure calculations that are compared with the ARPES data and we find reasonable agreement, confirming previous report\cite{Ding}. The measured intensity contours at higher binding energy ($\sim$ 500~meV) very closely resemble the features near $E_{\rm F}$ in BaFe$_2$As$_2$, which demonstrates that replacing Fe ions with Co ions causes a nearly rigid shift in the Fermi energy by adding two electrons per (f.u.)  for limited range of binding energies. We also report the presence of relatively flat bands in the proximity of $E_{\rm F}$, that are most likely responsible for peak in DOS ($E_{\rm F}$). This may be important for understanding the transport and thermodynamic properties of this material.

\section{Experimental and theoretical details}

Single crystals of BaCo$_2$As$_2$ were grown out of CoAs self-flux.\cite{Bishop10, Sefat09} The chemical compositions were checked by wavelength-dispersive x-ray spectroscopy (WDS) that revealed the desired stoichiometry of $1:2:2$. A Rietveld refinement of x-ray diffraction data for crushed crystals confirmed that BaCo$_2$As$_2$ crystallizes in the body-centered tetragonal ThCr$_2$Si$_2$ structure with lattice parameters $a=3.9536(4)$~\AA~and $c=12.641(2)$~\AA~at room temperature, in good agreement with literature values.\cite{Pfisterer}

We have performed high-resolution angle-resolved photoemission spectroscopy (ARPES) measurements using a Scienta R4000 electron analyzer at beamline 7.0.1 at the Advanced Light Source (ALS), Berkeley, California. All samples were cleaved {\it in situ}, yielding flat mirror-like surfaces in the {\it a-b} plane. All the ARPES data were collected in ultrahigh vacuum below $4\times10^{-11}$~mbar. The energy and momentum resolution were set to $\sim20$~meV and $\sim0.3^{\circ}$, respectively. We measured several samples, which yielded similar results for the Fermi surfaces. The $E_{\rm F}$ was determined from the measurements on a Au reference sample. 

%The FPLAPW method with the local density approximation (LDA) was used to calculate the theoretical FS and band dispersions. To obtain self-consistent charge density, we employed RMT*kmax=8.0  with muffin tin (MT) radii of 2.5, 2.1 ,2.1 a.u. for Ba, Co and As respectively. 828 k-points were selected in the irreducible Brillouin zone and calculations were iterated to reach the total energy convergence criterion which was 0.01 mRy/cell. As atom was relaxed to get theoretical position that was zAs=0.344101

The full-potential linearized augmented-plane-wave (FP-LAPW) method with the local density approximation\cite{Perdew92} was used to calculate the theoretical FS and band dispersions. To obtain self-consistent charge density, we employed $R_{\rm MT}\times{\it k}_{\rm max}=8.0$ with muffin tin (MT) radii of 2.5, 2.1, and 2.1~a.u. for Ba, Co, and As, respectively. 828~$k$-points were selected in the irreducible Brillouin zone and calculations were iterated to reach the total energy convergence criterion which was 0.01 mRy/cell. %\textcolor{red}{(definition?)}
The As atom {\it z}-axis position was relaxed to obtain a minimum total energy, yielding {\it z}$_{\rm As}=0.3441$. For the Fermi surface calculation, we divided the $-2\pi/a \le k_{\it x}$, $k_{\it y}\le 2\pi/a$ ranges of $k_{\it x}$, $k_{\it y}$ planes with different $k_{\it z}$ values into 200$\times$200 meshes. 

\section{Results and discussion}

\begin{figure}
\includegraphics[width=4in]{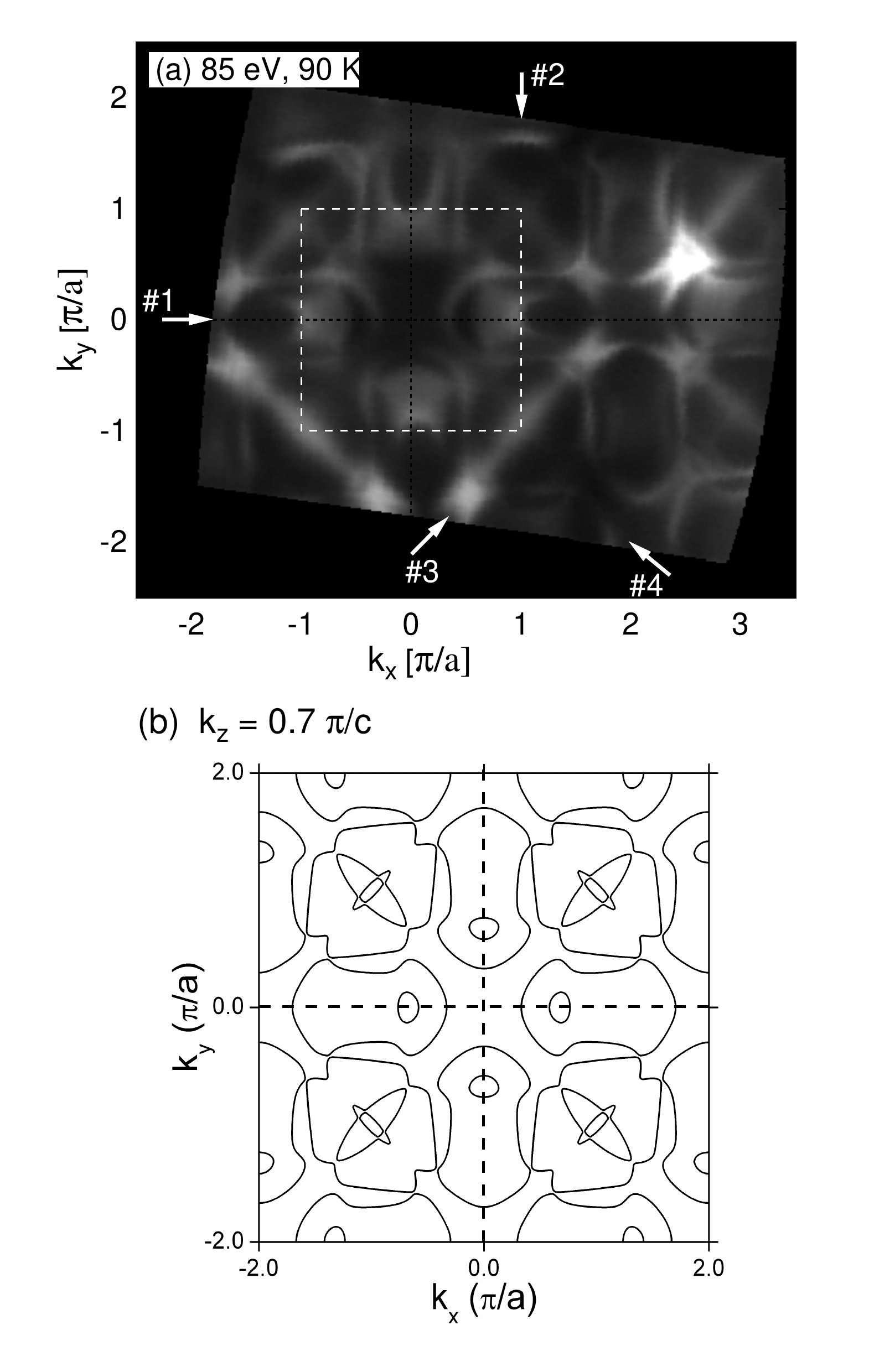}%55BaCo2As2_85eV90K_FS3
\caption{(color online) Comparison between the measured ARPES intensity map at $E_{\rm F}$ and the theoretical FP-LAPW calculations. The box (dashed line) marks the boundary of the first Brillouin zone. Arrows indicate the locations of cuts for which the data are plotted in Figs~3 and 4. (a) Experimental FS of BaCo$_2$As$_2$ at $T=90$~K measured with 85~eV photon energy. The intensity of the photoelectrons is integrated over $\pm$10~meV about  $E_{\rm F}$. (b) The calculated Fermi surface at $k_z=0.7\pi/c$.}
\label{fig1}
\end{figure}

\begin{figure}
\includegraphics[width=4in]{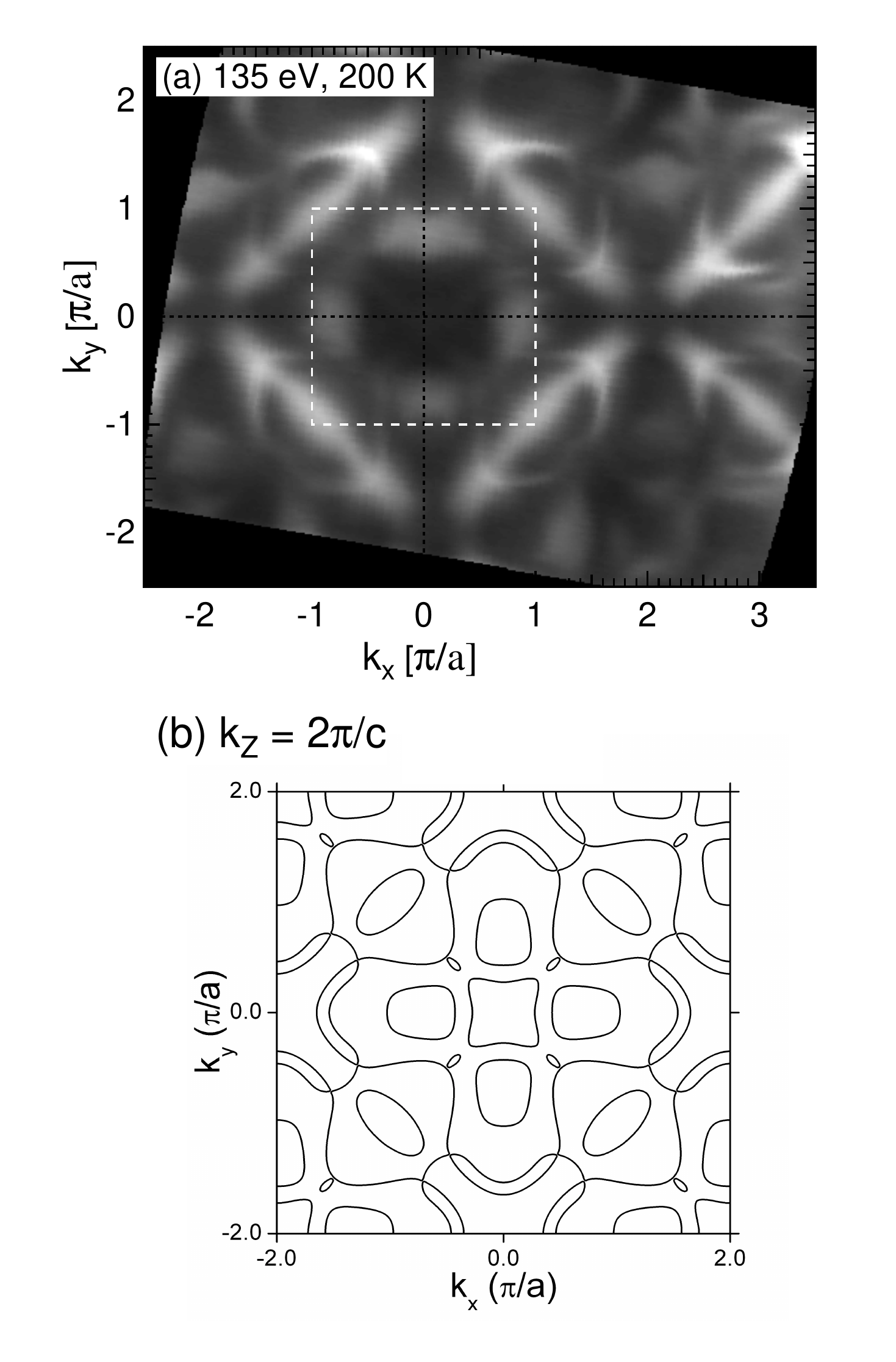}%54BaCo2As2_135eV200K_FS1
\caption{ (a) Fermi surface map of BaCo$_2$As$_2$ measured with 135~eV photon energy and at $T=200$~K. The intensity of the photoelectrons is integrated over $\pm$10~meV about $E_{\rm F}$. (b) The calculated FS at $k_z=2\pi/c$.}
\label{fig3}
\end{figure}

\begin{figure}
\includegraphics[width=6in]{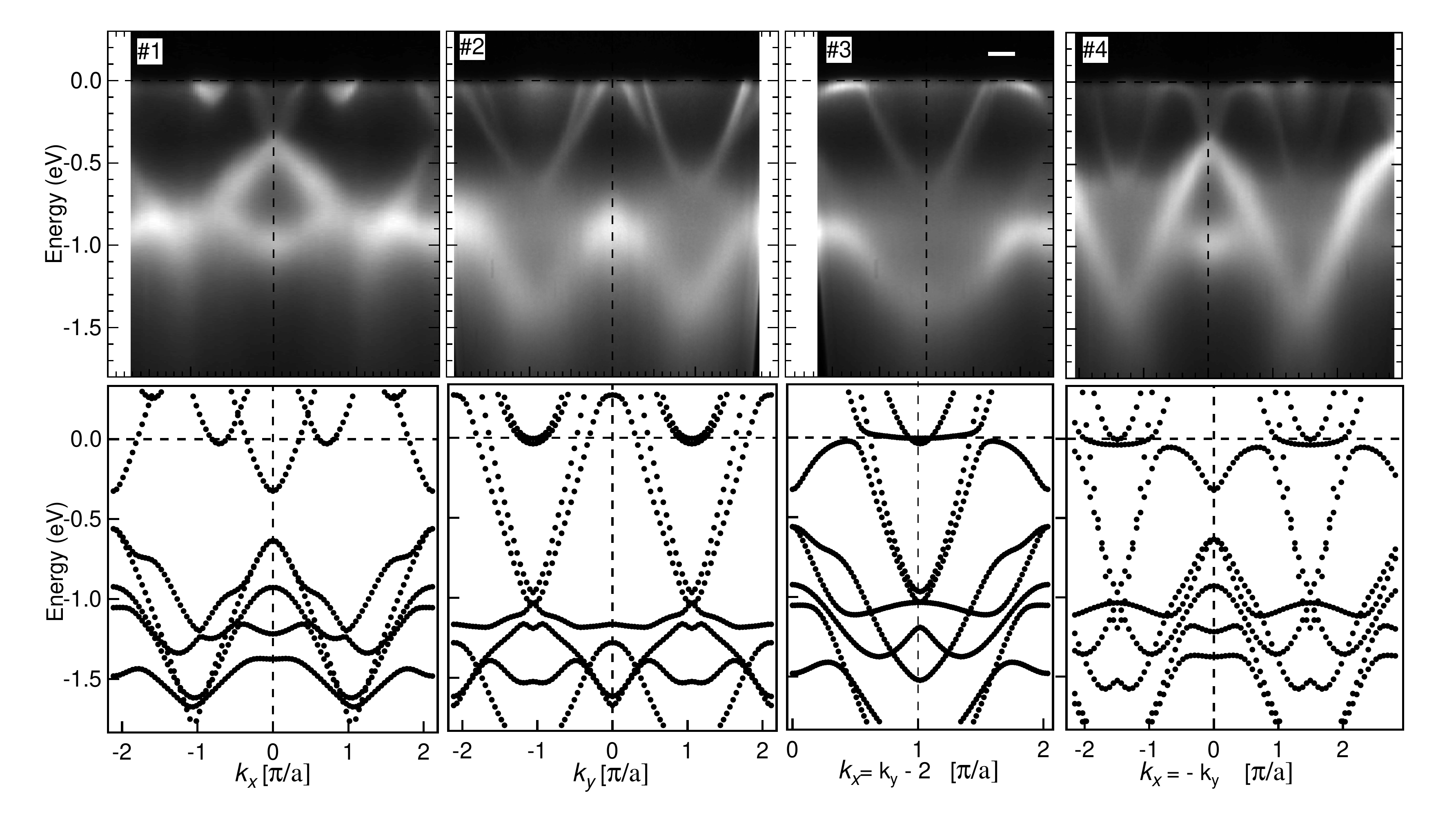}%55BaCo2As2_85eV90K_FS3
\caption{ (Upper panels) Experimental band dispersion data of BaCo$_2$As$_2$, measured with 85~eV photon energy and at $T=90$~K.  The locations of the cuts \#1, \#2, \#3 and \#4 in $k_x, k_y$ space are marked in Fig.~1(a). (Lower panels) Calculated band dispersions at $k_z=0$. The white bar at the top of cut \#3 shows the range of momenta for which EDCs are plotted in Fig. 4.}
\label{fig1}
\end{figure}

\begin{figure*}
\includegraphics[width=6in]{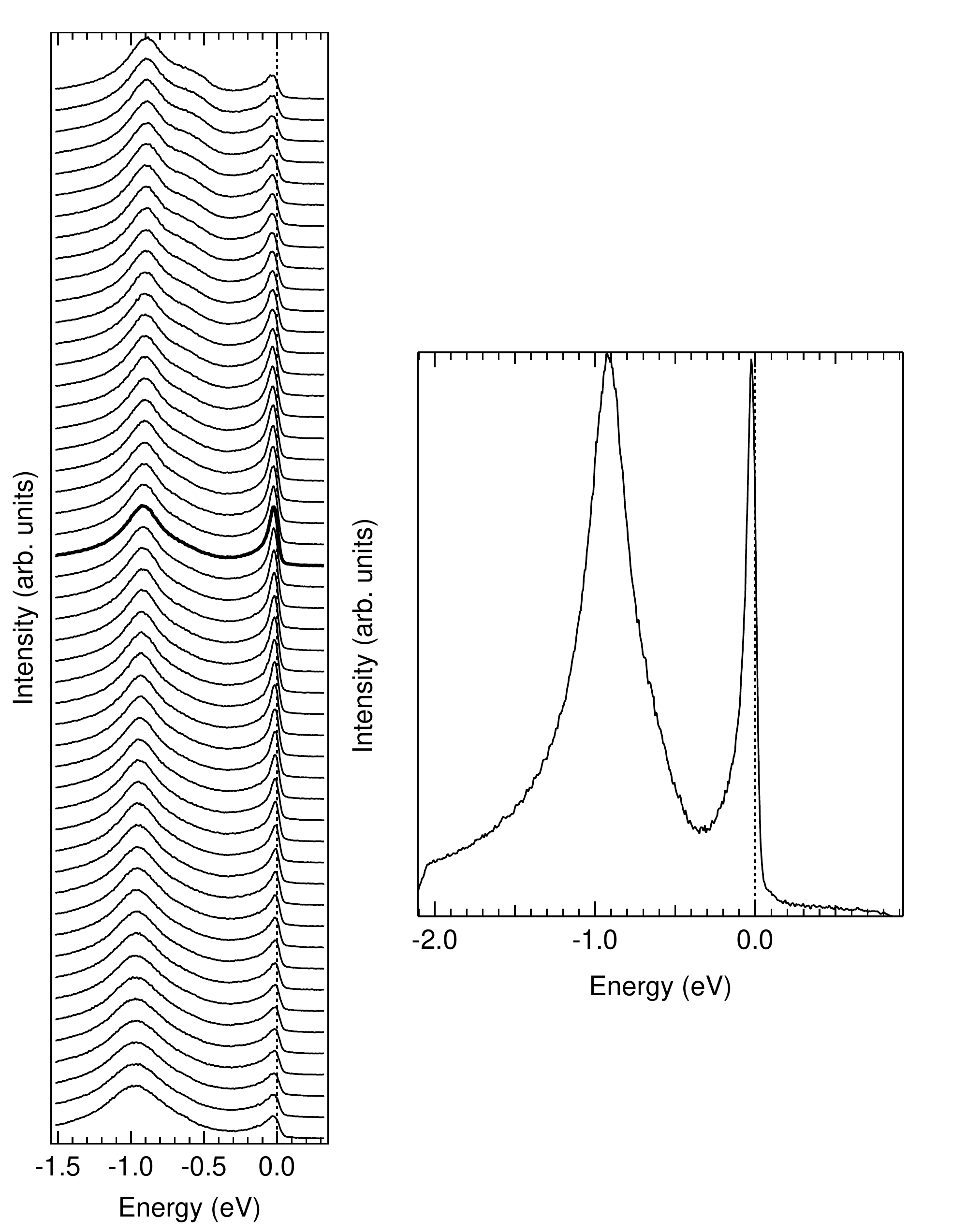}%55BaCo2As2_85eV90K_bands3
\caption{(Left panel) EDCs for small range of momenta in cut 3 (location and range indicated by white bar in cut \#3, Fig.~3). (Right panel) Single EDC marked by the bold curve in the left panel.}
\label{fig2}
\end{figure*}

\begin{figure*}
\includegraphics[width=6in]{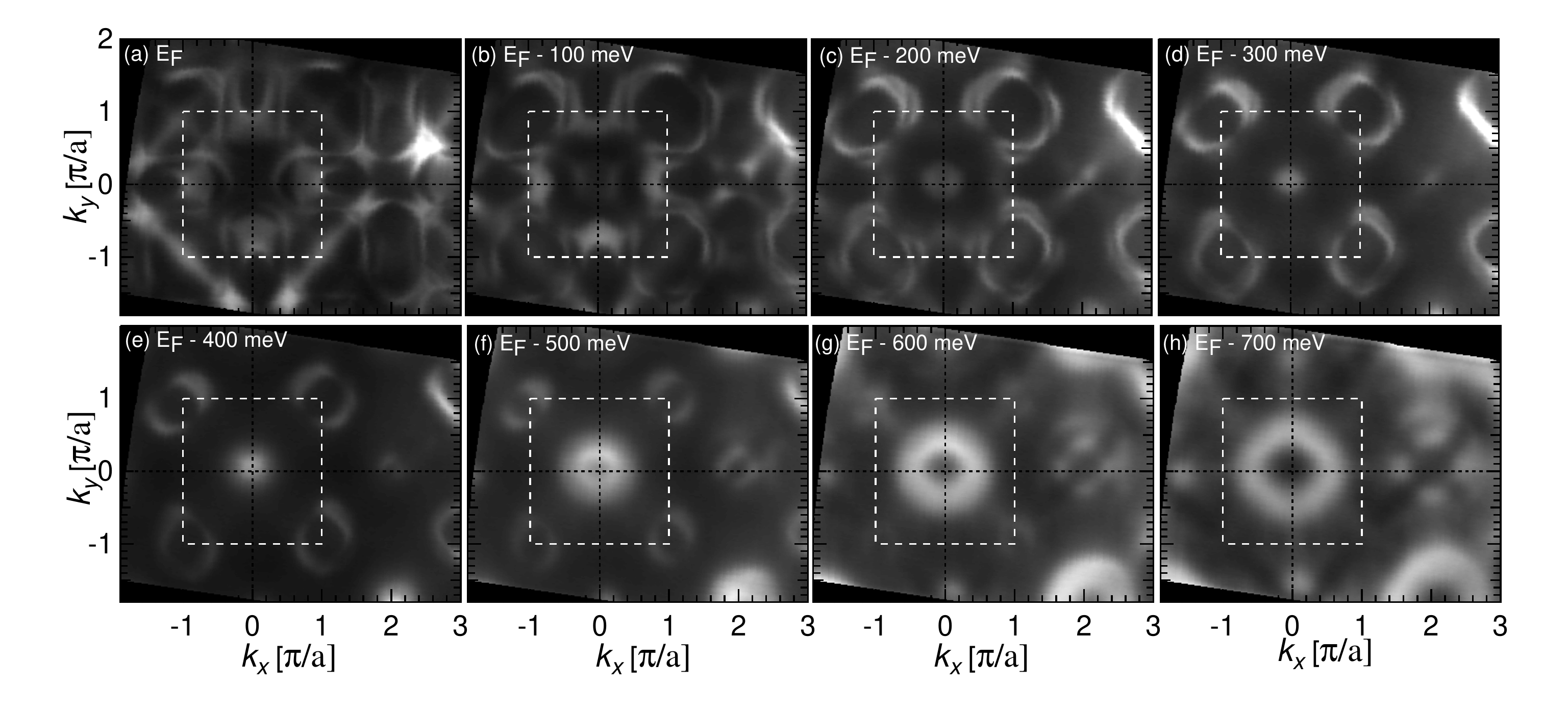}%55BaCo2As2_85eV90K_FS_energy
\caption{ Comparison of the FS maps of BaCo$_2$As$_2$ at different binding energies. These maps are measured with 85~eV photon energy and at $T=90$~K. }
\label{fig1}
\end{figure*}

\begin{figure*}
\includegraphics[width=6in]{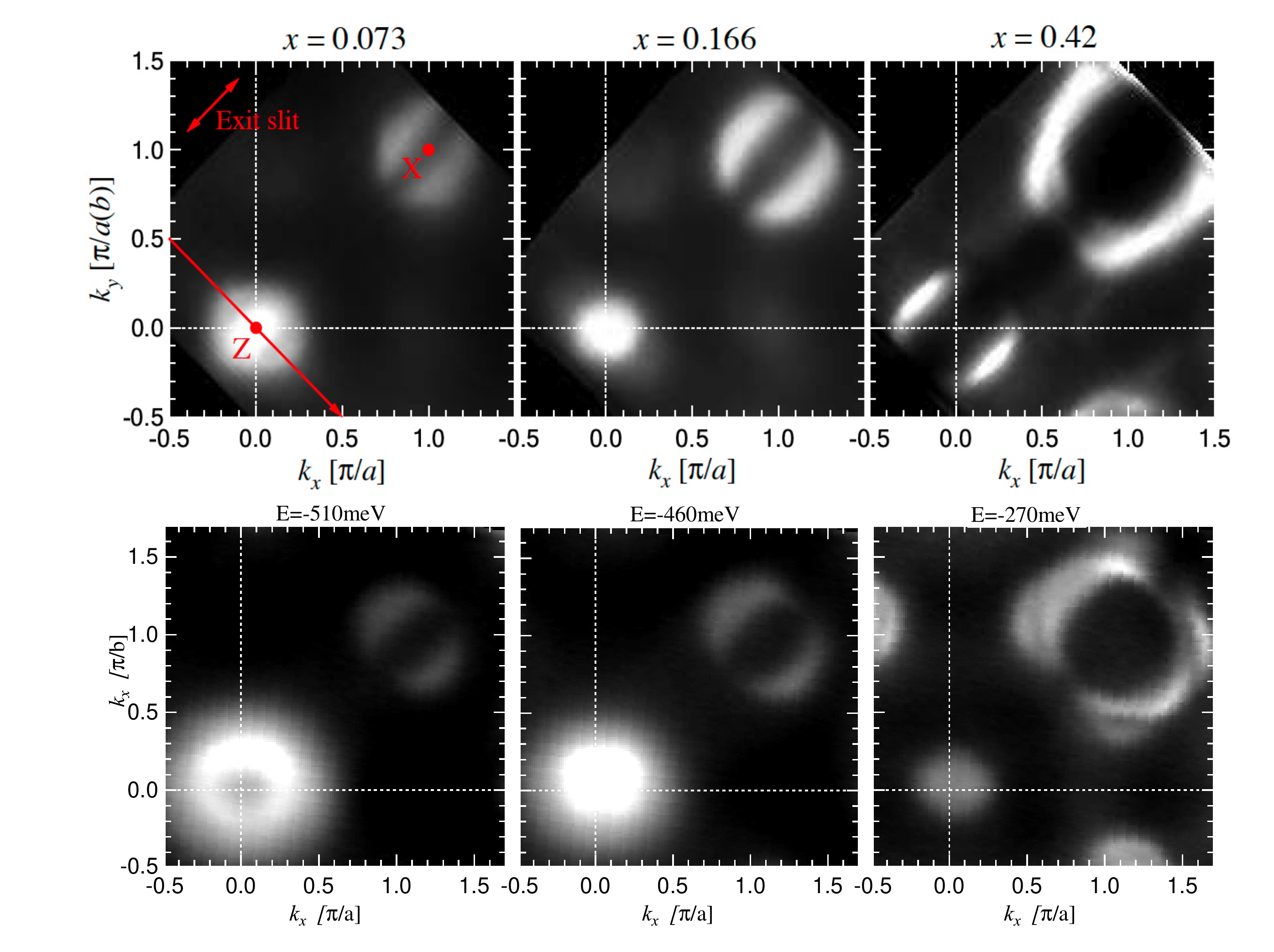}%55BaCo2As2_85eV90K_FS_energy
\caption{ (Top panel) Comparison of the intensity plots of Ba(Fe$_{1-x}$Co$_x$)$_2$As$_2$ at $E_{\rm F}$ (Fermi surface) for three values of $x$ (from Ref.~\onlinecite{LiuPRB11}). (Bottom panel)Constant binding energy intensity plots of BaCo$_2$As$_2$ (from the same data set as in Fig. 5)}
\label{fig1}
\end{figure*}

Figure 1(a) shows the photoemission intensity map at $E_{\rm F}$  of  BaCo$_2$As$_2$ measured with 85~eV photon energy  at temperature 90~K. The map is plotted as a function of {\it k}$_x$, {\it k}$_y$ wave vector components and the intensity is obtained by integrating the spectra within $\pm$10~meV of $E_{\rm F}$. The shapes of the FS sheets in BaCo$_2$As$_2$ are very different from the roughly circular FSs present in FeAs-based materials, which is not surprising, given that Co contributes one electron more than Fe. There is a significant difference in the shape of the intensity pattern between the first and second Brillouin zones, most likely due to a combination of matrix element effects and three dimensionality of the electronic structure.  Since ARPES cannot measure the magnitude of the ${\it z}$-component of the momentum, we use a comparison with calculations to determine the relation between $k_{\it z}$  and photon energy. In Fig.~1(b) we plot the calculated band structure that best matches  the data measured with 85 eV photons. The corresponding value of  $k_{\it z}$ in the calculations is 0.7$\pi/c$.  Figure 2(a) shows the intensity map at  $E_{\rm F}$  measured with 135~eV photons, while panel (b) shows the calculated FS for $k_{\it z}=2\pi/c$. The reasonable agreement in the overall shapes of the features indicates that we have made the correct assignment of $k_{\it z}$. 

%\begin{figure}
%\includegraphics[width=3.5in]{Fig4}%53_55_BaCo2As2_85eV_90K_200K
%\caption{(color online) The Fermi surface maps of BaCo$_2$As$_2$ measured with 85~eV photon energy and at (a) 90~K and (b) 200~K. Intensity of the photoelectrons is integrated over 20~meV about the $E_{\rm F}$.}
%\label{fig3}
%\end{figure}

The Fermi surfaces shown in Figs.~1 and 2 consist of several sheets. To examine their character, we have plotted in Fig.~3 (upper panels) the experimental band dispersions  along the four different momentum  cuts \#1, \#2, \#3 and \#4 shown in  Fig.~1(a). The band dispersion plot in cut \#1 is from ($-2\pi$, 0) to ($2\pi$, 0), which shows a small electron pocket at the center of the BZ ($\Gamma$ point) with Fermi momenta $k_{\rm F\it x}=\pm 0.3 \pi/a$. The bottom of this electron pocket is located  about 300~meV below $E_{\rm F}$ and it touches the top of a fully occupied band. An appearance of the electron pocket at $\Gamma$ is expected based on a doping study of Ba(Fe$_{1-x}$Co$_x$)$_2$As$_2$ (Ref.~\onlinecite{LiuPRB11}) since Co  provides one extra electron and shifts  $E_{\rm F}$ upwards in energy. In addition,  two smaller pockets are observed at Fermi momenta $k_{\rm F\it x}=\pm 0.9 \pi/a$. The band dispersion plot in cut \#2 is from $(k_{\it x}, k_{\it y})=(2\pi, \pi)$ to $(-2\pi, -\pi)$  and shows larger electron pockets at the corner of the first BZ at ${\rm X}=(\pi,\pi)$ with $k_{\rm F\it y}=\pm 0.8 \pi/a$. Cut \#3 from $(k_{\it x}, k_{\it y})= (0, -2\pi)$ to $(2\pi, 0)$  shows a large electron pocket that has  Fermi momenta similar to the previous one. Finally, we plot  cut \#4, from $(-2\pi, 2\pi)$ to $(2\pi, -2\pi)$, which again shows a small electron pocket at $\Gamma$ and large electron pockets at the $X$ points. In the lower panels, we show the corresponding calculated band dispersions, which are in reasonable qualitative agreement with the experimental data. We also note the presence of a rather flat portions of the bands in close proximity to  $E_{\rm F}$ visible in both the data and calculations for cuts \#3 and \#4. These likely lead to an increase of the density of states at $E_{\rm F}$ and therefore should have consequences for transport and thermodynamical properties. To further demonstrate this, we plot in Fig.~4 (left panel) the energy distribution curves (EDCs) in the range of momentum where this flat band is located (marked by a white bar at the top of cut \#3 in Fig. 3). The EDC peaks are nearly dispersionless and located in very close proximity to $E_{\rm F}$. The intensity of these peaks is very high, approaching the intensity of the valence band, as shown in the right panel of Fig.~4.

In Fig.~5, we compare the intensity maps of BaCo$_2$As$_2$ measured at different binding energies. It is seen that the electron pocket at the center of the BZ shrinks to a point at about 400~meV below $E_{\rm F}$ [Fig.~5(e)]. Below this energy a hole-like contour is present that increases in size with increasing binding energy. At about 500~meV below $E_{\rm F}$ the size of this hole contour is comparable to the hole FS sheet observed in BaFe$_2$As$_2$. Similarly, the electron pocket at the corner of the BZ shrinks in size with increasing binding energy and reaches the size comparable to the FeAs-based material at about 500~meV binding energy. In fact the intensity plot in Fig.~5(f) looks very much like the FS of unsubstituted BaFe$_2$As$_2$. This indicates that substitution of Co for Fe mostly shifts the chemical potential without significantly changing the band dispersions. Thus, fully substituting the Fe by Co corresponds to an upward shift of the chemical potential by $\sim$500~meV. In Fig.~6, we compare the Fermi surface of Ba(Fe$_{1-x}$Co$_x$)$_2$As$_2$ (adopted from Ref.~\onlinecite{LiuPRB11}) with constant energy intensity plots at high binding energy of BaCo$_2$As$_2$ samples. The values of binding energies were chosen to match roughly the size of the contours of the electron bands in the upper and lower panels (Fig.~6), respectively. We note that  the plots at high binding energy of BaCo$_2$As$_2$ quite well match the shape and size of the Fermi surface of Ba(Fe$_{1-x}$Co$_x$)$_2$As$_2$, implying that a rigid band shift is a good approximation for  low lying bands. However, at lower binding energies in BaCo$_2$As$_2$  and higher cobalt concentrations in Ba(Fe$_{1-x}$Co$_x$)$_2$As$_2$, the shape and size of the contour at the zone center are quite different, despite the good match at the zone corners. This signifies that the rigid band shift breaks down above -460 meV.

\section{Conclusions}

We presented an angle-resolved photoemission study of the Fermi surface and band structure of BaCo$_2$As$_2$  and compared these with theoretical calculations carried out using the FP-LAPW method. Reasonable qualitative agreement in the shape of the FS and band dispersion plots between the experiment and calculations was found. The Fermi surfaces consist of electron pockets both at the center and corner of the BZ. The size of these pockets is very different prohibiting interband nesting. The measurements performed at higher binding energies ($\sim$500 meV) reveal intensity contours that are almost identical to the Fermi surface of BaFe$_2$As$_2$. This agreement indicates that full substitution of Fe by Co results in a chemical potential increase of about 500 meV without significant renormalization in the band dispersions.

\section{Acknowledgments}

We thank Aaron Bostwick and Eli Rotenberg for excellent support at the ALS and Abhishek Pandey for helpful discussions.  Research supported by the U.S. Department of Energy, Office of Basic Energy Sciences, Division of Materials Sciences and Engineering.  Ames Laboratory is operated for the U.S. Department of Energy by Iowa State University under Contract No. DE-AC02-07CH11358. The Advanced Light Source is supported by the Office of Basic Energy Sciences, U. S. Department of Energy under Contract No. DE-AC02-05CH11231.

\end{document}